\documentclass[creativecommons]{eptcs}
\providecommand{\event}{E. Formenti (Ed.): AUTOMATA \& JAC 2012}
\usepackage[T1]{fontenc}
\usepackage[utf8]{inputenc}
\usepackage{amsmath,amssymb,amscd,amsfonts,graphicx,tikz}
\usepackage[format = hang]{caption}
\usetikzlibrary{patterns}
\newcommand\tx{\text}
\newcommand\A{\mathcal{A}}
\newcommand\N{\mathbb N}
\newcommand\Z{\mathbb Z}

\newcommand\R{\mathbb R}
\newcommand\U{\mathbb U}
\newcommand\conf{\A^\Z}
\newcommand\s{\sigma}
\newcommand\ds{\displaystyle}

\newcommand\mes{\mathcal{M}(\conf)}
\newcommand\inv{\mathcal{M}_{\s}(\conf)}

\newcommand\Ber{\mathcal Ber}
\newcommand\Mix{\mathcal Mix}

\newcommand{\lsem}{[}
\newcommand{\rsem}{]}
\newcommand{\prv}{\noindent\noindent{\textit{Proof.}} }
\newcommand{\qed}{\hfill$\square$}
\newtheorem{dfe}{Definition}
\newtheorem{lem}{Lemma}
\newtheorem{thm}{Theorem}
\newtheorem{prop}{Proposition}
\newtheorem{cor}{Corollary}
\begin{document}
\title{Entry times in automata with simple defect dynamics.}
\author{Benjamin Hellouin de Menibus \qquad\qquad Mathieu Sablik
\institute{Laboratoire d'Analyse, Topologie, Probabilités (LATP)\\
Université d'Aix-Marseille\\
Marseille, France}
\email{benjamin.hellouin@gmail.com \qquad\qquad sablik@latp.univ-mrs.fr}
}
\def\titlerunning{Entry times in automata with simple defect dynamics}
\def\authorrunning{B. Hellouin de Menibus, M. Sablik}
\maketitle{}

\begin{abstract}
  In this paper, we consider a simple cellular automaton with two particles of different speeds that annihilate on contact. Following a previous work by K\r urka et al., we study the asymptotic distribution, starting from a random configuration, of the waiting time before a particle crosses the central column after time n. Drawing a parallel between the behaviour of this automata on a random initial configuration and a certain random walk, we approximate this walk using a Brownian motion, and we obtain explicit results for a wide class of initial measures and other automata with similar dynamics.
\end{abstract}

\section{Introduction}
Self-organization in cellular automata is the emergence of structures when one iterates a cellular automaton on a random initial configuration. For some cellular automata, self-organization takes the form of the emergence and the persistence of homogeneous regions separated by boundaries which propagate and sometimes collide over time like particles. These particles and their dynamics under the action of the CA have been studied empirically (\cite{Boc}, \cite{HanCru}), and Pivato proposed a general formalism to describe this phenomenon \cite{Piv2}. Under some assumptions on the dynamics of particles, it is possible to get information on the asymptotic behaviour of the automaton. See for example \cite{Elo}, \cite{Vanish}, or \cite{Nous}, where we proved the following: when particles have constant speeds and destructive interactions, and under some assumptions on the initial measure, the probability for particles to appear in a central cylinder tends to 0 as time tends to infinity except possibly for one particular speed.\\

\begin{figure}[h]
\begin{center}
\begin{tabular}{cc}
$(-1,0)$-gliders automaton&3-state cyclic automaton\\
\includegraphics[width=5.9cm]{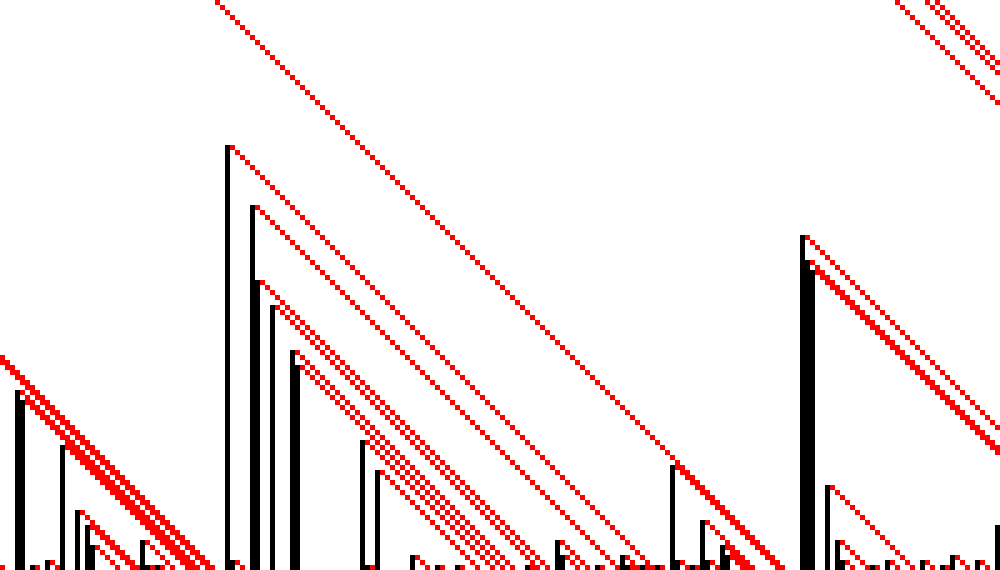}&\includegraphics[width=5.9cm]{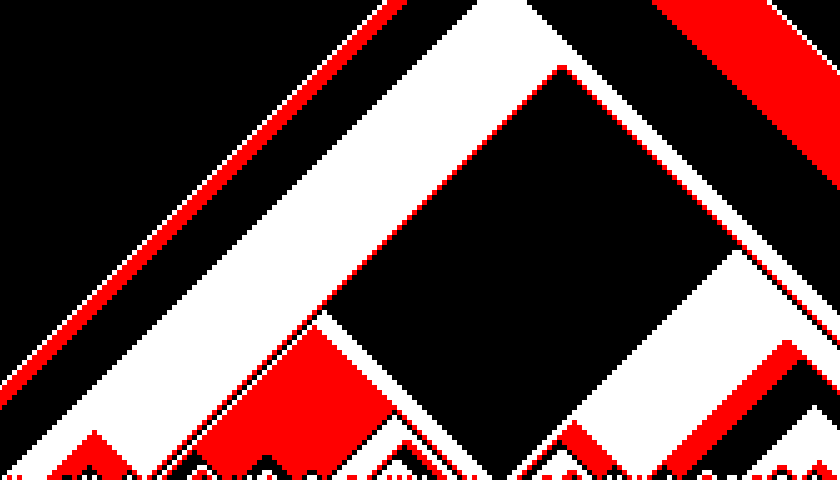}\\
Traffic automaton (Wolfram rule \#184)&One-sided captive automaton\\
\includegraphics[width=5.9cm]{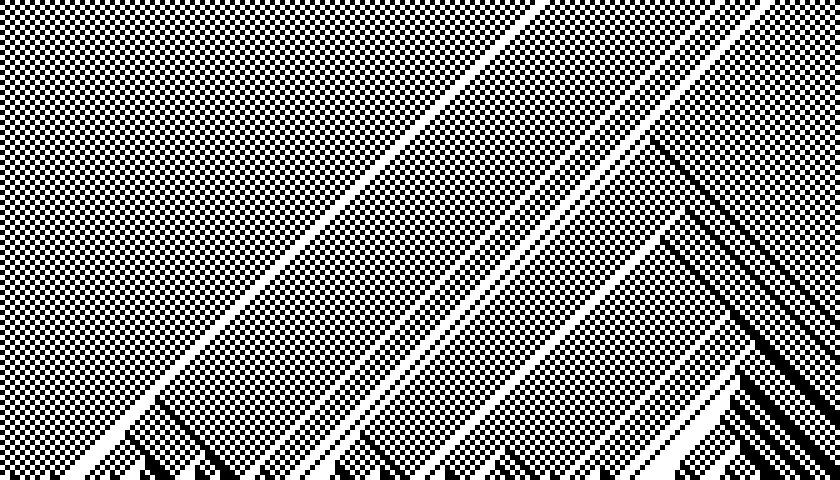}&\includegraphics[width=5.9cm]{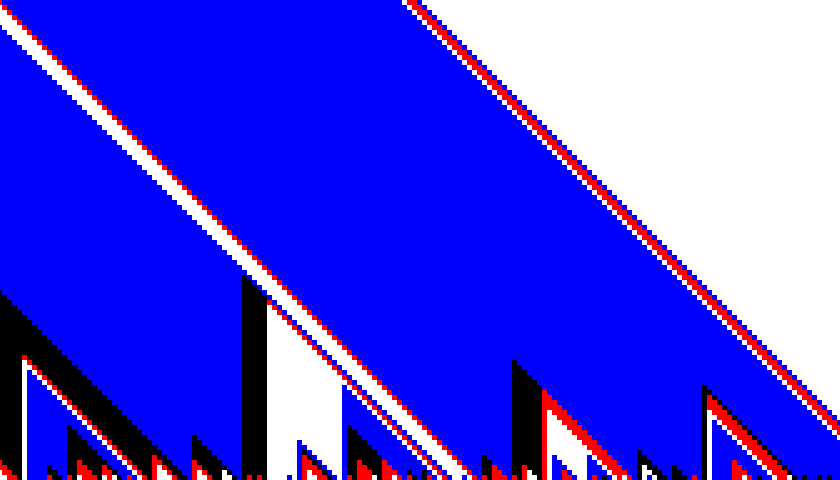}
\end{tabular}
\end{center}
\captionsetup{hangindent = -.8cm}
\caption{Some automata with simple defect dynamics.\newline{}
In all diagrams, time goes from bottom to top.}
 \label{intro}
\end{figure}

For some automata with simple dynamics, this kind of results can be refined with a quantitative approach: that is, to determine the asymptotic distribution of some random variable related to the particles. In \cite{Entry}, K\r urka, Formenti and Dennunzio considered $T_n(a)$, the entry time after time $n$ on an initial configuration $a$, which is the waiting time before a particle appears in a given position after time $n$. They restricted their study to a gliders automaton, which is a cellular automaton on 3 states: a background state and two particles evolving at speeds 0 and -1 that annihilate on contact. Thus, we have one entry time for each type of particle ($T_n^+(a)$ and $T_n^-(a)$). When the initial configuration is drawn according to the Bernoulli measure of parameters $(\frac 12, 0, \frac 12)$, which means that each cell contains, independently, a particle of each type with probability $\frac 12$, they proved that 
\[\forall x\in\R^+,~\mu\left(\frac {T_n^-(a)}n \leq x\right)\underset{n\to\infty}{\longrightarrow}\frac 2\pi \arctan\sqrt x.\]
They also called to develop formal tools in order to be able to handle more complex automata, starting with the $(-1,1)$ symmetric case.\\

In section 3, we extend in some sense this result to allow arbitrary values for the speeds $v_-$ and $v_+$, and relax the conditions on the initial measure to some $\alpha$-mixing conditions. Then, when $v_-< 0$ and $v_+\geq 0$, we have:
\[\forall x\in\R^+,~\mu\left(\frac{T_n^-(a)}n\leq x\right) \underset{n\to\infty}{\longrightarrow} \frac 2\pi\arctan\left(\sqrt{\frac{-v_-x}{v_+-v_-+v_+x}}\right),\]
and symetrically if we exchange $+$ and $-$. The proof relies on the fact that the behaviour of gliders automata can be characterized by some random walk process;  his general idea was introduced by K\r urka \& Maass in $\cite{KurMaa}$ and was already used in \cite{Entry}. In our case, a particle appearing in a position corresponds to a minima between two concurrent random walks. Under $\alpha$-mixing conditions, we rescale this process and approximate it with a Brownian motion. Thus we obtain the explicit asymptotic distribution of entry times.\\

Furthermore, this result can be extended to other automata with similar behaviour, such as those in Fig.~\ref{intro}, by factorizing them onto a gliders automaton. This point is discussed in section 4.\\
\section{Definitions}
\subsection{Cellular automata}

Let $\A$ be a finite alphabet. We consider the spaces $\A^*=\cup_{n\in\N}\A^{\lsem0,n\rsem}$ of finite \textbf{words} and $\conf$ of bi-infinite \textbf{configurations}. We note $|u|$ the \textbf{length} of any word $u$. For $a\in\conf$, define its {\bf subwords} $a_{[x,y]}=u_x\dots u_y\in\A^{y-x+1}$ for $x,y\in\Z$. $\conf$ is compact in the product topology, and the \textbf{cylinders} $[u]_m = \{a\in \conf: a_{[m,m+|u|-1]} = u\}$ for $u\in \A^*$ and $m\in\Z$ are clopen sets and form a base for this topology.\\

The \textbf{shift} function $\s: \conf \to \conf$ is defined by $(\s(a))_v = a_{v+1}$ for all $a\in\conf$ and $v\in\Z$. A \textbf{cellular automaton} or CA is a continuous function $F: \conf\to\conf$ which commutes with $\s$. Equivalently, $F$ is a CA iff there exists a \textbf{local rule} $f: \A^{\U} \rightarrow \A$, where $\U$ is a neighbourhood of the origin, such that $F(a)_z = f(a_{z+\U})$. We study the action of $F$ on $\conf$, and especially the sequence $(F^n(a))_{n\in\N}$ for some initial configuration $a$. For a (partial) graphical representation, we associate a color with each state and represent some finite window of $(F^n(a)_k)_{k\in\Z, n\in\N}$ (the {\bf space-time diagram}).

\begin{dfe}
 Let $v_-,v_+ \in \Z$ such that $v_-<v_+$. The \textbf{$(v_-,v_+)$-gliders automaton} (or GA) is the CA of neigbourhood $[-|v_+|, -|v_-|]$ defined on the alphabet $\A = \{-1, 0, +1\}$ by the local rule:
\[f(a_{-r}\dots a_r) = \left\{\begin{array}{cl}+1&\tx{if }a_{-v_+}=+1\tx{ and }\forall N\leq -v_-,\sum_{t=-v_++1}^{N} a_t\geq 0\\-1&\tx{if }a_{-v_-}=-1\tx{ and }\forall N\geq -v_+,\sum_{t=N}^{-v_--1} a_t\leq 0\\0&\tx{otherwise.}\end{array}\right.\]
\end{dfe}
In the space-time diagrams of gliders automata, we adopt the convention $\square = 0, \blacksquare = +1, \textcolor{red}{\blacksquare} = -1$.
\begin{figure}[h]
\begin{center}
 \includegraphics[scale=0.5]{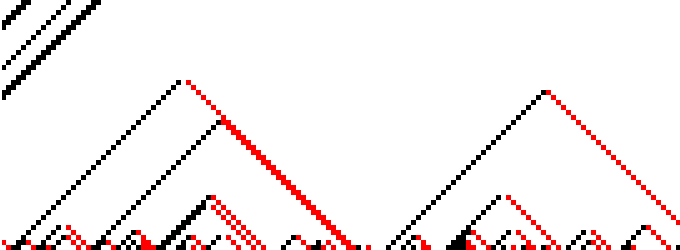}
 \caption{Space-time diagram of the $(-1,1)$-gliders automaton on a random initial configuration.}
\end{center}
\end{figure}

\subsection{Measures on $\conf$}
 Let $A$ be a finite alphabet. We define $\mes$ the set of probability measures on the borelians of $\conf$. In all the following, we write $\forall_\mu a$ for ``for $\mu$-almost all $a$''.

\begin{dfe} 
 Let $\pi : \conf \to \mathcal B^\mathbb Z$ a measurable application. It induces an action $\pi_{\ast}: \mes \to \mathcal M(\mathcal B^\mathbb Z)$ by defining $\pi_{\ast}\mu(U) = \mu(\pi^{-1} U)$ for any borelian $U$. We write $\pi\mu$ instead of $\pi_{\ast}\mu$ to simplify notations.
\end{dfe}

\textbf{$\s$-action:} We note $\inv$ the {\bf $\s$-invariant probability measures} on the borelians of $\conf$, i.e. the measures such that $\s\mu= \mu$. In this case we note $\mu([u])$ for $\mu([u]_0)$. For any $k\in\Z$, let $\gamma_k$ be the projection $\gamma_k(a) = a_k$. By $\s$-invariance, if $a$ is drawn according to $\mu$, then $(a_k)_{k\in\Z}$ is a sequence of stationary, non necessarily independent random variables, each drawn according to $\gamma_0\mu$.\\

\noindent{\textbf{Examples:}}
\vspace{-0.1cm}
\begin{description}
 \item[Dirac measure] Let $a\in\conf$ be a periodic configuration of minimal period $n$. We define $\delta_a(U)=\frac kn$, where $k = \big|\{0\leq i\leq n-1\ :\ \s^i(a)\in U\}\big|$.
 \item[Bernoulli measure] Let $(p_{a})_{a\in\A}$ be a sequence satisfying $\sum_{a\in\A}p_a=1$. Then, define $\mu$ as $\mu([u])=p_{u_0}p_{u_1}\cdots p_{u_{|u|-1}}$.
 \item[2-step Markov measure] Let $(p_{ij})_{i,j\in\A}$ be a matrix satisfying $\sum_j p_{ij} = 1$ for all $i$, and let $(\mu_i)$ an eigenvector associated with the eigenvalue 1 (which is unique if the matrix is irreducible). The associated Markov measure is defined as $\mu([u]) = \mu_{u_0}p_{u_0u_1}\cdots p_{u_{|u|-2}u_{|u|-1}}$. Define similarly an $n$-step Markov measure.
\end{description}

We introduce the notion of \textbf{$\alpha$-mixing} for a measure $\mu\in\inv$. Let $\mathcal R_n^{+\infty}$, resp. $\mathcal R_{-\infty}^0$, be the $\s$-algebra generated by the cylinders $\{[u]_n\ |\ u\in\A^*\}$, resp. $\{[u]_{-k}\ |\ k\in\mathbb N, u\in\A^k\}$. The {\bf $\alpha$-mixing coefficients} of $\mu$ are \[\alpha_\mu(n) = \sup \{|\mu(A\cap B) - \mu(A)\mu(B)|: A\in \mathcal R_{-\infty}^0, B\in \mathcal R_n^\infty\}\]

\textbf{$F$-action:} Consider the sequence $(F^k\mu)_{k\in\N}$. Putting on $\inv$ the weak-* topology, we consider the set of limit points $\Gamma_F(\mu)$ of this sequence.\\

The $\mu$-limit set $\Lambda_F(\mu) = \bigcup_{\nu \in \Gamma_F(\mu)}supp(\nu)$ is of particular interest for self-organization, as discussed in \cite{KurMaa}. Indeed, consider words that appear arbitrarily far in space-time diagrams, i.e., such that $F^n\mu([u]) \not\to 0$ ({\bf $\mu$-persistent} words). Then, one can show that a configuration appears in the $\mu$-limit set iff all its subwords are $\mu$-persistent.\\

In all the following, $\A = \{-1,0,1\}$. We consider two particular subclasses of $\inv$:
\begin{itemize}
 \item $\Ber$ the set of Bernoulli measures of parameters $(p,1-2p,p)$ for some $0<p\leq \frac 12$;
 \item $\Mix$ the set of measures satisfying:
    \begin{itemize}
    \item $\mathbb E(\gamma_0\mu) = 0$;
    \item $\s_\mu^2 = \mathbb E(\gamma_0\mu^2) + \sum_{k=1}^\infty \mathbb E(\gamma_0 \mu \cdot \gamma_k \mu)>0$ (asymptotic variance);
    \item $\exists \varepsilon>0,\sum_{n\geq 0}\alpha_\mu (n)^{\frac 12 - \varepsilon} < \infty$.
    \end{itemize}
\end{itemize}

In particular, $\Ber \subset \Mix$. The hypotheses for $\Mix$ are chosen so that the large-scale behaviour of the partial sums $M_a(k)$, defined by $M_a(0) = 0$ and $\forall k\in\Z, M_a(k+1)-M_a(k) = a_k$, can be approximated by a Brownian motion. This invariance principle is the core of our proofs. 

\begin{dfe}
 A {\bf Brownian motion} (or {\bf Wiener process}) $B$ of mean $\mu$ and variance $\sigma^2$ is a a continuous time stochastic process taking values in $\R$ such that:
\begin{itemize}
\renewcommand{\labelitemi}{\labelitemii}
 \item $B(0) = 0$,
 \item $t\mapsto B(t)$ is almost surely continuous,
 \item $B(t_2) - B(t_1)$ follow the normal law of mean 0 and variance $(t_2 - t_1)\s^2$;
 \item For $t_1<t_2\leq t'_1<t'_2$, increments $B(t_2) - B(t_1)$ and $B(t'_2) - B(t'_1)$ are independent.
\end{itemize}
\end{dfe}

See \cite{Brown} for a general introduction to Brownian motion.

\begin{thm}[\cite{Invariance}]\label{Brown} Let $\mu \in \Mix$. Then, $\s_\mu^2<\infty$ and it is possible to construct a sequence $(Z_i)_{i\in\Z}$ of centered Gaussian variables of variance $\s_\mu^2$ such that \[\forall \alpha>0, \forall_\mu a\in \conf, \sup_{-n\leq k\leq n}\left|M_a(k) - \sum_{i=0}^k Z_i\right| = o\left(n^{1/2}\right).\]In other words, if we rescale the process: \[S_a: \begin{array}{l}\R\to \R\\ t\mapsto (t-|t|)M_a(|t|)+(|t|+1-t)M_a(|t|+1)\end{array}\quad\tx{ and }\quad S_a^k: t\mapsto\frac{S_a(kt)}{\sqrt k}.\]
Then, if $x<y\in\mathbb R$ are any fixed constants, we can construct a Brownian process $B_a$ of parameters $(0,\s>0)$ on $[x,y]$ such that:
\[\forall_\mu a\in \conf, \sup_{[x,y]}\big|S^n_a(t) - B_a(t)\big| \underset{n\to\infty}{\longrightarrow}0\]
\end{thm}
For a survey of invariant principles under different assumptions, see \cite{MerlRio}.\\

\section{Main result}

\subsection{Entry times}
The main result of \cite{Nous} implies that, for any $\sigma$-ergodic initial measure $\mu$ (this includes Bernoulli and 2-step Markov measures; see op.cit. for exact definitions), $\Lambda_F(\mu)$ contains at most one kind of particle, which one depending on whether $\mu([+1]) > \mu([-1])$ or the opposite. If $\mu([+1]) = \mu([-1])$, for example when $\mu \in \Mix$, $\Lambda_F(\mu)$ only contains the particleless configuration. This implies that $F^n\mu$ converges to the Dirac measure on this configuration, which means that the probability of seeing a particle in any given column tends to 0 as $t\to\infty$. 

\begin{dfe}[Entry times]
 Let $v_-< 0 \leq v_+\in\Z$ and $a \in\conf$. We define: \[T_n^-(a) = \min\{k\in\N\ |\ \exists i\in[0,|v_-|-1],F^{k+n}(a)_i=-1\},\] with $T_n^-(a) = \infty$ if this set is empty. This is the \textbf{entry time} of $a$ into the set $\{b\in\conf\ |\ \exists i\in[0,|v_-|-1], b_i=-1\}$ after time $n$ at position 0. We define $T_n^+(a)$ in a similar manner.
\end{dfe}

\begin{figure}
\begin{center}
 \includegraphics[scale = 0.8]{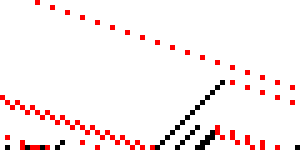}
  \hspace{-9.31cm}
\begin{tikzpicture}
\draw[blue,thick] (0,0) -- (0,4.2);
\draw[blue,thick] (0.42,0) -- (0.42,4.2);
\draw[blue,thick] (0,1) rectangle (0.42,1.14);
\draw[blue,thick] (0,1) -- (0.42,1.14) (0.42,1) -- (0,1.14);
\filldraw[blue!15!white] (0.02,1.16) rectangle (0.40,3.09);
\draw[<->, thick] (-0.2,1.14) -- (-0.2,3.1);
\draw (-0.7, 2.12) node {\small $T_n^-(a)$};
\draw[<->,thick] (-0.2,1) -- (-0.2,0);
\draw (-0.5, 0.5) node {\small $n$};
\draw (-4,-0.01) -- (5,-0.01) (-4,0.13) -- (5,0.13);
\draw (-4.25,0.2) node {\small $a$};
\end{tikzpicture}
\end{center}
\caption{An entry time for the (-3,1)-gliders automaton.}\label{ent}
\end{figure}
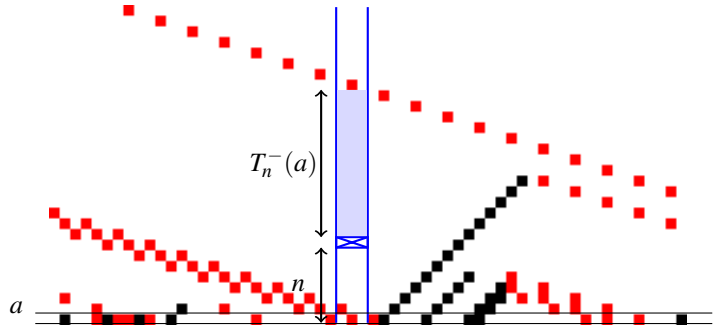
The size of the considered window is such that any particle ``passing through'' the position 0 will appear in this window exactly once (see Fig.~\ref{ent}). Of course entry times for particles of speed 0 make no sense. From now on, we will only consider $T^-$ for simplicity, all the results being valid for $T^+$.\\

As a direct consequence of Birkhoff's ergodic theorem, we see that when $\mu([-1]) > \mu([+1])$:
\begin{itemize}
 \item $\mu(T^+_n(a)=\infty) \underset{n\to\infty}{\longrightarrow} 1$;
 \item $\mu\left(\frac {T_n^-(a)}n \leq x\right)\underset{n\to\infty}{\longrightarrow} 1$.
\end{itemize}

Kurka \& al. proved the following result:
\begin{thm}[cf. \cite{Entry}]
 For the $(-1,0)$-GA (``Asymmetric gliders''), if the initial measure $\mu$ is the Bernoulli one of parameters $(\frac 12,0,\frac 12)$, then we have asymptotically:\[\forall x\in\R^+,~\mu\left(\frac {T_n^-(a)}n \leq x\right)\underset{n\to\infty}{\longrightarrow}\frac 2\pi \arctan\sqrt x.\]
\end{thm}
In the same article, they conjectured that this result could be extended to any initial Bernoulli measure of parameters $(p,1-2p,p)$ by replacing the right-hand term by $\frac 2\pi \arctan\sqrt{2px}$.
\begin{thm}[Main result]\label{entry}
 For any $(v_-,v_+)$-GA with $v_-< 0$ and $v_+\geq 0$, with the initial measure $\mu\in\Mix$, we have asymptotically: 
\[\forall x\in\R^+,~\mu\left(\frac{T_n^-(a)}n\leq x\right) \underset{n\to\infty}{\longrightarrow} \frac 2\pi\arctan\left(\sqrt{\frac{-v_-x}{v_+-v_-+v_+x}}\right).\] 
\end{thm}
Notice that this limit is independent of the variance of $\pi_0\mu$, disproving the conjecture when $\mu\in\Ber$.

\subsection{Technical lemmas}

\begin{lem}\label{lem1}
 $\forall j\in\Z, \forall n\geq 1$,
\begin{align*}
 M_{F(a)}(j) < \underset{\{j+1,\dots, j+n\}}{\min} M_{F(a)} \Leftrightarrow M_{a}(j-v_+) < \underset{\{j+1-v_+,\dots, j+n-v_-\}}{\min} M_a,\\
 M_{F(a)}(j) < \underset{\{j-n,\dots, j-1\}}{\min} M_{F(a)} \Leftrightarrow M_{a}(j-v_-) < \underset{\{j-n-v_+,\dots, j-1-v_-\}}{\min} M_a.
\end{align*}
and those inequalities still hold if we replace < by $\leq$.
\end{lem}

\prv We prove those four inequalities by induction on $n$. At each step, we will prove only the first inequality with $<$, the other cases being symmetric.\\

\noindent{\textit{Base case.}} 
\begin{align*}
M_{F(a)}(j) < M_{F(a)}(j+1) &\Leftrightarrow F(a)_j = +1\\
                            &\Leftrightarrow a_{j-v_+} = +1 \tx{ and } \forall N\leq -v_-,\sum_{t=-v_++1}^N a_{j+t}\geq 0\\
                            &\Leftrightarrow  M_{a}(j-v_+) < \underset{\{j+1-v_+,\dots, j+1-v_-\}}{\min} M_a
\end{align*}
\textit{Induction.} Assume the four inequalities hold for some $n$. We distinguish two cases:
\begin{itemize}
\item if $F(a)_{j} \neq 1$, then $a_{j-v_+} \neq +1$, or $a_{j-v_+} = +1$ and $\exists N\leq v_-, \sum_{t=j-v_+}^{j+N}a_t < 0$. In any case,
\[M_{F(a)}(j) \geq \underset{\{j+1,\dots, j+n+1\}}{\min} M_{F(a)} \tx{  and  } M_{a}(j-v_+) \geq \underset{\{j+1-v_+,\dots, j+n-v_-\}}{\min} M_a\]
and so the inequality holds.
 \item if $F(a)_{j} = 1$, then $a_{j-v_+} = 1$. We again distinguish two cases:
    \begin{itemize}
     \item If $M_{F(a)}(j+1) \leq \underset{\{j+2,\dots, j+n+1\}}{\min} M_{F(a)}$, then $M_{a}(j-v_++1) \leq \underset{\{j-v_++2,\dots, j+n-v_-+1\}}{\min} M_a$
by induction hypothesis, and since $M_{F(a)}(j) =  M_{F(a)}(j+1)-1$ and $M_{a}(j-v_+)=M_{a}(j-v_++1)-1$, the inequality holds.\\
    \item Otherwise, $M_{F(a)}(j+1) > \underset{\{j+2,\dots, j+n+1\}}{\min} M_{F(a)}$ and $M_{a}(j-v_++1) > \underset{\{j+2-v_+,\dots, j+n-v_-\}}{\min} M_a$ by induction hypothesis, and the inequality holds for the same reason.\qed
    \end{itemize} 
\end{itemize}

\begin{figure}
\begin{center}
\begin{tabular}{cc}
\hspace{-0.2cm}\includegraphics{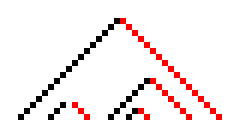}&\\
\includegraphics{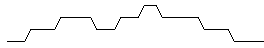} & \vspace{-5.95cm}\\
\hspace{-0.35cm}
\begin{tikzpicture}
\draw[blue, thick] (-0.1,3.25) -- (-3.35,0) -- (-3.35, -1.55);
\draw[blue, thick] (0.1,3.25) -- (3.35,0) -- (3.35,-1.55) ;
\draw (-3.7,-0.75) node {$S_a$};
\draw[red, thick] (-3.35, -1.55) -- (3.35,-1.55);
\draw (-3.35,-1.8) node {\small $j-k+1$};
\draw (3.35,-1.8) node {\small $j+k$};
\draw[<->, thick] (-4, 0) -- (5,0);
\draw (-4.2,0) node {$a$};
\draw[->] (-3.5, 0) -- (-3.5,3.25);
\draw (-3.7,1.625) node {$k$};
\draw (0,-0.1) -- (0,0.1);
\draw (0, -0.3) node {\small $j$};
\filldraw[blue] (-0.1,3.25) rectangle (0.1, 3.45);
\draw (-0.4,3.7) node {\small $F^k(a)_j$};
\end{tikzpicture}&
\end{tabular}
\end{center}
\caption{Illustration of lemma \ref{lem2}. A strict minimum is reached on $j-k+1$.}

\label{figbiz}
\end{figure}

\begin{lem}\label{lem2}$\forall j\in\Z, \forall k\geq 0,$
\begin{align*}F^{k}(a)_j= -1 \Leftrightarrow M_a(j-v_-k+1) < \underset{\{j-v_+k,\dots, j-v_-k\}}{\min} M_a\\
  F^{k}(a)_j= +1 \Leftrightarrow M_a(j-v_+k) < \underset{{\{j-v_+k+1,\dots, j-v_-k+1\}}}{\min} M_a
\end{align*}
This is illustrated in Fig.~\ref{figbiz}, where $M_a$ is replaced by its piecewise affine interpolation $S_a$.
\end{lem}

\prv By induction on $k$, proving only the first equality at each step:\\

\noindent{\textit{Base case.}} Obviously, $M_a(j+1)<M_a(j) \Leftrightarrow a_j=-1$.\\

\noindent{\textit{Induction.}} Now suppose those equalities hold for a given rank $k$. By induction hypothesis, $F^{k+1}(a)_j= -1 \Leftrightarrow M_{F(a)}(j-v_-k+1) < \underset{\{j-v_+k,\dots, j-v_-k\}}{\min} M_{F(a)}$ and we conclude by lemma \ref{lem1}.\qed

\subsection{Proof of the theorem} 

For any $a\in \conf$, the lemma 2 on the column 0 gives:
\begin{align*}
T_n^-(a)& =\min\left\{k\geq 0\ |\ \exists j\in [0,-v_-[~, M_a(-v_-(n+k)+j+1) < \min_{\{-v_+(n+k)+j,\dots,-v_-(n+k)-j\}}M_a\right\}\\ 
& =\min\left\{k\geq 0\ |\ \exists j\in [0,-v_-[~, M_a(-v_-(n+k)+j+1) < \min_{\{-v_+(n+k)+j,\dots,-v_-n\}}M_a\right\}
\end{align*}
We keep notations from Thm.~\ref{Brown}: $S_a$ is the piecewise affine interpolation of $M_a$ and $S_a^n$ is a rescaling of $S_a$. Note that when the previous condition is attained on $M_a(x)$, then it is attained for $S_a(y)$ as soon as $y>x-1$, and reciprocally. Thus:
\begin{align*}
T_n^-(a)&=\inf\left\{t\geq 0\ |\ \exists j\in [0,-v_-[~, S_a(-v_-(n+t)+j+2) < \min_{[-v_+(n+t)+j+1,-v_-n]}S_a\right\}\\
&=\left\lceil\inf\left\{t\geq 0\ |\ S_a(-v_-(n+t)+2) < \min_{[-v_+(n+t)+1,-v_-n]}S_a\right\}\right\rceil\\
&=\left\lceil\inf\left\{t\geq 0\ |\ S_a^n\left(-v_-\left(1+\frac 2n\right)+\frac1n\right) < \min_{[-v_+(1+\frac tn)+\frac1n,-v_-]}S_a^n\right\}\right\rceil\\
&=\left\lceil n\cdot\inf\left\{t\geq 0\ |\ S_a^n\left(-v_-(1+t)+\frac 2n\right) < \min_{[-v_+(1+t)+\frac1n,-v_-]}S_a^n\right\}\right\rceil
\end{align*}
And finally, since $S_a^n$ is $\sqrt n$-Lipschitz and $\forall x,n\in\R\times\N$, $x \leq \frac{\lceil nx\rceil}n \leq x+\frac1n$:
\begin{align*}
\mu\left(\min_{[-v_-,-v_-(1+x)]}S_a^n +\frac3{\sqrt n}< \min_{[-v_+(1+x),-v_-]}S_a^n\right) \leq \mu\left(\frac{T_n^-(a)}n \leq x\right) \\ \mu\left(\frac{T_n^-(a)}n \leq x\right) \leq \mu\left(\min_{[-v_-,-v_-(1+x)]}S_a^n -\frac4{\sqrt n}< \min_{[-v_+(1+x),-v_-]}S_a^n\right)\tag{1}\label{eq1}
\end{align*}

Using Thm.~\ref{Brown}, we can construct a Brownian motion $B_a$ such that $S_a^n \to B_a$ for the $||.||_\infty$ norm on $L^\infty([-v_+(1+x),-v_-(1+x)])$. By symmetry, $B_a^l(t) = B_a(-v_--t)-B_a(-v_-)$ and $B_a^r(t) = B_a(-v_-+t)-B_a(-v_-)$ are two independent Brownian motion satisfying $B_a^l(0) = B_a^r(0) = 0$. Consequently, for any $\varepsilon >0$ and $n$ large enough:

\begin{align*}\mu\left(\min_{[-v_-,-v_-(1+x)]}S_a^n -\varepsilon< \min_{[-v_+(1+x),-v_-]}S_a^n\right) &\leq \mu\left(\min_{[-v_-,-v_-(1+x)]}B_a-2\varepsilon<\min_{[-v_+(1+x),-v_-]}B_a\right)\\
&\leq \mu\left(\min_{[0,-v_-x]}B_a^l-2\varepsilon<\min_{[0,-v_-+v_+(1+x)]}B^r_a\right)\tag{2}\label{eq2}
\end{align*}

and similarly for the left-hand term of (\ref{eq1}). For a brownian motion $B$ and any $b>0$, we have by rescaling $\ds\mu\left(\min_{[0,b]} B\geq m\right) = \mu\left(\min_{[0,1]} B\geq \frac m{\sqrt b}\right)$. Furthermore, since $B^l_a$ and $B^r_a$ are independent, so are $\ds\min_{[0,1]} B^l_a$ and $\ds\min_{[0,1]} B^r_a$. That means that for any $y,z>0$:

\begin{align*}
\mu\left(\min_{[0,y]}B^l_a< \min_{[0,z]} B^r_a\right) &=\int_{-\infty}^0\int_{-\infty}^01_{\{\sqrt y\cdot m_1\leq \sqrt z\cdot m_2\}}d\mathbb P_{\{\min B^r_a\}}(m_2)d\mathbb P_{\{\min B^l_a\}}(m_1)\\
&\underset{(i)}=\frac 4{2\pi}\int_{-\infty}^0\int_{-\infty}^{\frac {\sqrt{z}\cdot m_2}{\sqrt y}} e^{\frac{-m_1^2}2}e^{\frac {-m_2^2}2} dm_1dm_2\\
&\underset{(ii)}=\frac 2{\pi}\int_\pi^{\pi+\arctan\left(\sqrt {\frac yz}\right)}\int_0^{+\infty} re^{\frac{-r^2}2} drd\theta\\
&=\frac 2{\pi}\arctan\left(\sqrt {\frac yz}\right)\label{eq3}\tag{3}
\end{align*}

(i) by using the law of the minimum of a Brownian motion (see \cite{Brown}), (ii) by passing in polar variables. For $\varepsilon>0$, a similar calculation gives:

\begin{align*}\left|\mu\left(\min_{[0,y]}B^l_a-2\varepsilon< \min_{[0,z]}B^r_a\right)-\mu\left(\min_{[0,y]}B^l_a< \min_{[0,z]}B^r_a\right)\right| &\leq \frac 4{2\pi}\int_{-\infty}^0\int_{\frac {\sqrt z\cdot m_2}{\sqrt y}}^{\frac {\sqrt z\cdot m_2+2\varepsilon}{\sqrt y}} e^{\frac{-m_1^2}2}e^{\frac {-m_2^2}2} dm_1dm_2\\
&\leq\frac {8\varepsilon}{2\pi\sqrt y}\int_{-\infty}^0 e^{\frac{-ym_2^2}{2z}}e^{\frac {-m_2^2}2} dm_2\\
&\underset{\varepsilon \to 0}{\longrightarrow}0\label{eq4}\tag{4}
\end{align*}

And similarly for the left-hand term. Combining (\ref{eq1}), (\ref{eq2}), (\ref{eq3}) and (\ref{eq4}), the theorem follows.\qed

\section{Extension to other automata}

\begin{dfe}Let $F_1, F_2$ be two CA on $\conf$ and $\mathcal B^{\Z}$, respectively. We say that $F_1$ \textbf{factorizes onto} $F_2$ if there exists a \textbf{factor} $\pi: \conf\to \mathcal B^{\Z}$, i.e. a continuous transformation that commutes with $\s$, such that $\pi\circ F_1 = F_2 \circ\pi$.\end{dfe}

Similarly to a CA, a factor is entirely defined by a neighbourhood $\mathbb U$ and a local rule $p: \A^{\mathbb U}\to\mathcal B$. See~\cite{LindMar} for a description of the role of factors in symbolic dynamics.\\

In this section, we will extend the Thm.~\ref{entry} to automata that factorize onto a gliders automaton, starting by showing how to find such a factor. In the examples given in Fig.~\ref{intro}, a similar behaviour is observed: starting from a random configuration, strips constituted of periodic patterns appear and persist, and the boundaries between these strips behave as particles of constant speed. Intuitively, these automata exhibit the same behaviour as a gliders automaton, if we see the regular patterns as the background and the boundaries as particles.

\begin{dfe}
 Let $\A$ a finite alphabet. A {\bf subshift of finite type} (SFT) $\Sigma\subset\conf$ is a set of configurations defined by a set of finite {\bf forbidden patterns} $U \subset \A^r$, where $r>0$ is the {\bf order} of the SFT. Precisely,
\[a \in \Sigma \Leftrightarrow \forall i\in\mathbb Z, a_{[i,i+r-1]}\not\in U.\] 
\end{dfe}

The set of {\bf defects} of $a\in\conf$ with regard to $\Sigma$ is $\mathbb D_\Sigma(a) = \{ i\in\Z : a_{[i,i+r-1]} \in U\}$. Even when $\Sigma$ is $F$-invariant (which means that the strips are persistent), there is no general relationship between defects of $a$ and defects of $F(a)$. However, some automata have defects (relatively to some SFT, the choice being usually obvious) behave as particles of constant speed with only two possible speeds $v_+$ and $v_-$; in other words, we can separate the forbidden patterns into two sets $U^+$ and $U^-$, with corresponding defects $\mathbb D^+_\Sigma(a)$ and $\mathbb D^-_\Sigma(a)$, and we have:
\[\begin{array}{rl}
i \in \mathbb D^+_\Sigma(F(a))\Leftrightarrow& i-v_+\in\mathbb D^+_\Sigma(a)\\&\tx{ and }\forall N\leq -v_-, \left|[-v_++1,N]\cap \mathbb D^+_\Sigma(a)\right| \geq \left|[-v_++1,N]\cap \mathbb D^-_\Sigma(a)\right|\\
i \in \mathbb D^-_\Sigma(F(a))\Leftrightarrow& i-v_-\in\mathbb D^-_\Sigma(a)\\&\tx{ and }\forall N\leq -v_-, \left|[N,-v_--1]\cap \mathbb D^-_\Sigma(a)\right| \geq \left|[N,-v_--1]\cap \mathbb D^+_\Sigma(a)\right|   
\end{array}\]
That is, the defects behave as the particles of the gliders automaton. Now it is easy to see that such an automaton factorizes onto the $(v_-,v_+)$-GA with the factor $\pi : \conf \to \{-1,0,1\}^\mathbb Z$ of neighbourhood $[-r,r]$, where $r$ is the order of $\Sigma$ and $\pi$ is defined by the local rule $p: \A^r\to\{-1,0,1\}$ defined as:
\[p(u) = \left\{\begin{array}l+1 \text{ if }u \in U^+;\\-1 \text{ if }u \in U^-;\\0 \text{ if }u \not\in U.\end{array}\right.\]

\begin{figure}[h]
\begin{center}
\includegraphics[width = 7cm]{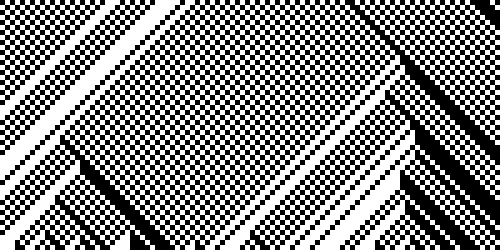}$\begin{array}{c}\to\vspace{3cm} \end{array}$\includegraphics[width = 7cm]{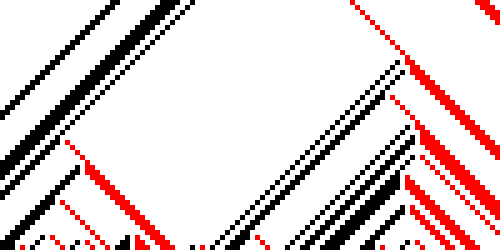}\vspace{-1.2cm}
\caption{Factor $\pi$: $\quad \square\hspace{-0.04cm}\square \to \blacksquare \quad\quad \blacksquare\hspace{-0.04cm}\blacksquare \to \textcolor{red}{\blacksquare} \quad\quad 
\square\hspace{-0.04cm}\blacksquare, \blacksquare\hspace{-0.04cm}\square \to \square.$}
\label{fig}
\end{center}
\end{figure}
\noindent{\textbf{Example:}} 
\begin{description}
\item[Traffic automaton:] Let $\A = \{0,1\}$ and $F$ be the CA of neighbourhood $\{-1,0,1\}$ defined by the local rule: \[f(u_{-1},u_0,u_1) = \left\{\begin{array}{ll}1& \text{if  }u_{-1} = 1\text{ and }u_0 = 0\\1& \text{if  }u_0=1\text{ and }u_1=1,\\ 0& \text{otherwise.}\end{array}\right.\] The behaviour of $F$ can be observed in Fig.~\ref{fig} with the convention $0 = \square, 1 = \blacksquare$. It is apparent that the relevant SFT corresponds to the set of forbidden patterns $U^+ = \{\square\square\}, U^- = \{\blacksquare\blacksquare\}$ (``checkerboard SFT'') and the corresponding defects have the dynamics of a $(-1,1)$-GA. The induced factor is indicated in Fig.~\ref{fig}.
\end{description}

In order to extend the Thm.~\ref{entry} to such CA with initial measure $\mu$, we must first ensure that $\pi\mu\in\Mix$. 

\begin{prop}
 Let $\pi: \conf \to \conf$ be a factor, $\mu\in\inv$ and $k>0$ any real such that $\sum_{n\geq 0}\alpha_\mu (n)^k < \infty$. Then, $\sum_{n\geq 0}\alpha_{\pi\mu} (n)^k < \infty$. 
\end{prop}

\prv We keep the notations from the definition of $\alpha_\mu(n)$. Since $\pi$ is continuous, there exists $r>0$ such that $\mathbb U \subset [-r,r]$ and $\pi(a)_k$ only depends on $a_{[k-r,k+r]}$. Then, $\pi^{-1}\mathcal R_{-\infty}^0 \subset \mathcal R_{-\infty}^r$ and $\pi^{-1}\mathcal R_n^\infty \subset \mathcal R_{n-r}^{+\infty}$. By $\s$-invariance, we have for all $n$ $\alpha_{\pi(\mu)}(n) < \alpha_\mu(n-2r)$, and the theorem follows. \qed\\

This is true in particular if $\mu$ is a Bernoulli measure or a 2-step Markov measure associated with a irreducible, aperiodic matrix. Hence, in that case, we only have to prove that $\mu$ weighs evenly the sets of particles $-1$ and $+1$, and that the corresponding asymptotic variance is not zero. Under those assumptions, we can extend the previous theorem with the forbidden patterns playing the role of the particles.

\begin{cor}
 Let $F$ be a CA on the alphabet $\A$ and $\mu\in\inv$. Suppose that $F$ factorizes onto a $(v_-,v_+)$-GA via a factor $\pi$, defined on a neighbourhood $[-r,r]$ by the local rule $p$, so that $\pi\mu\in\Mix$. 

Then, Thm.~\ref{entry} holds if we replace ``$a_k = -1$'' by ``$p(a_{[k,k+r-1]})=-1$'', and similarly for $+1$.
\end{cor}

\begin{figure}[h]
 \begin{center}
\includegraphics[width = 4.9cm]{cycle3cut.png}\hspace{0.5cm}\includegraphics[width = 4.9cm]{captifcut.png}\hspace{0.5cm}\includegraphics[width = 4.9cm]{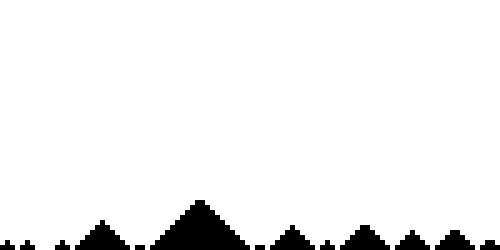}
 \end{center}
\caption{The 3-state cyclic CA, a one-sided captive CA and the product CA.}
\end{figure}
\noindent{\textbf{Examples:}} (In all the following, we use the convention $\square = 0, \blacksquare = 1, \textcolor{red}{\blacksquare} = 2, \textcolor{blue}{\blacksquare} = 3 $.)
\begin{description}
\item[Traffic automaton:] Consider the factor defined earlier. If $\mu$ is a measure such that $\pi\mu\in\Mix$, then Thm.~\ref{entry} applies. For example, this is true for the 2-step Markov measure defined by the matrix $\left(\begin{matrix}p&1-p\\1-p&p\end{matrix}\right)$ and the eigenvector $\left(\begin{matrix}1/2\\ 1/2\end{matrix}\right)$ with $p>0$. A particular case is the Bernoulli measure of parameters $(\frac 12,\frac 12)$.
 \item[Cyclic automaton:] Let $\A = \Z/3\Z$ and $F$ the CA of neighbourhood $\{-1,0,1\}$ defined by the local rule \[f(u_{-1},u_0,u_1) = \left\{\begin{array}{ll} u_0+1&\text{if }u_{-1} = u_0+1\text{ or }u_1=u_0+1,\\ u_0&\text{otherwise.}\end{array}\right.\]The relevant SFT is the SFT of order 2 obtained by forbidding $U^+ = \{\blacksquare \square, \square \textcolor{red}{\blacksquare},\textcolor{red}{\blacksquare}\blacksquare\}$, $U^-$ being symmetric (``monochromatic SFT''). The corresponding defects obviously behave as particles of speed +1 or -1 and $F$ factorizes onto the $(-1,1)$-GA. If $\mu$ is such that $\pi\mu\in\Mix$, then Thm.~\ref{entry} applies. This is true in particular when $\mu$ is any 2-step Markov measure defined by a matrix $(p_{ij})_{1\leq i,j\leq 3}$ satisfying $p_{01}+p_{12}+p_{20} = p_{10}+p_{21}+p_{02}$, all of these values being $>0$, with $(\mu_i)_{1\leq i\leq 3}$ its only eigenvector. This includes any nondegenerate Bernoulli measure.
 \item[One-sided captive automata:] Let $F$ be a CA of neighbourhood $\{0,1\}$ on any alphabet, such that $f(u_{-1},u_0) \in \{u_{-1},u_0\}$. We consider the monochromatic SFT, that is, the SFT of order 2 obtained by forbidding words with two different letters. For $a\neq b\in\A$, if $f(a,b) = a$, we consider that $ab\in U^+$; otherwise, $ab\in U^-$. In this way, $F$ factorizes onto the $(-1,0)$-GA, and if $\pi\mu\in\Mix$, then Thm.~\ref{entry} applies.\\

Notice that this class of automata contain the identity ($\forall a,b, f(a,b) = b$) and the shift $\s$ ($\forall a,b, f(a,b) = a$). However, since we have in each case $U^+ = \emptyset$ or $U^- = \emptyset$, it is impossible to find a measure that weighs evenly each kind of particle, and so $\pi\mu$ cannot belong in $\Mix$.
\end{description}

\noindent{\textbf{Counter-example:}}
\begin{description}
\item [Product automaton:] Let $\A = \Z/2\Z$ and $F$ be the CA of neighbourhood $\{-1,0,1\}$ defined by the local rule $f(a_{-1},a_0,a_{1}) = a_{-1}\cdot a_0\cdot a_{1}$. The relevant SFT corresponds to the forbidden patterns $U^+ = \{\blacksquare\square\}, U^- = \{\square\blacksquare\}$. Then, $F$ factorizes onto the $(-1,1)$-GA. If $\mu$ is a Bernoulli measure, then $\pi\mu$ satisfies all conditions of $\Mix$ except that $\s_\mu = 0$; indeed, we can check that for $\pi\mu$-almost all configurations, the particles $+1$ and $-1$ altern. Hence, only one particle can cross any given column after time 0, and therefore $\mu\left(\frac{T_n^-(a)}n\leq x\right) \underset{n\to\infty}{\longrightarrow} 0$.
\end{description}

\section{Conclusion}
Even though we showed that the asymptotic distributions of entry times are known for some class of cellular automata and a large class of measures, this covers only very specific dynamics. It is not known how these results extend for more than 2 particles and/or other kind of particle interaction. In particular, there is no obvious stochastic process characterizing the behaviour of such automata that would play the role of $M_a$ in our proofs.

\section*{Acknoledgments}
We would like to thank Xavier Bressaud for his helpful insights and an anonymous reviewer for its numerous corrections and suggestions. All space-time diagrams were made with Sage (\cite{Sage}).


%
\end{document}